# Three distinctive temperatures of normal-to-superconductive phase transition detected by the unique <u>SFCO method</u>: *their crucial role for proper identification of the nature of superconductivity*

**Samvel G. Gevorgyan**[1,2]

[1] Faculty of Physics, Yerevan State University, 1 Alex Manoogian str., Yerevan, 0025, ARMENIA
[2] Institute for Physical Researches, National Academy of Sciences, Ashtarak-2, 0203, ARMENIA

**Abstract**
An improved *'LC-resonator'* method (a low-power stable tunnel diode oscillator with non-traditional single-layer flat pick-up coil) is an excellent *MHz*-range instrument with which one may study weakly expressed peculiarities of the superconductive (**SC**) state in tiny plate-like high-$T_c$ superconductive (**HTSC**) materials. It enables to detect changes *~1 pH* of the *HTSC* film's magnetic inductance (changes *~1-3 Å* of the *SC*-magnetic penetration depth, $\lambda$) with *~6 orders* relative resolution. Due to detecting coil's flat design, relatively low operation frequency, so high relative resolution and an ability of this unique technique to operate also in high magnetic fields the method has advantages over all others. They become crucial at non-destructive studies in thin flat *HTSC* materials with a small signal – especially near the normal-to-superconductive (**N/S**) phase transition (at start of formation of electron-pairs). Just due to such advantages a very fine *'paramagnetic'* peculiarity of the *SC*-transition was detected by this method in YBaCuO film recently, which precedes well-known *Meissner* expel. The method enabled to detect also a little absorption of a power by the *HTSC* film upon its transition, with a peak located after the onset point of the *Meissner* push-out - close to its center. Extracted from so complicated fluctuation temperature-region these subtle effects (1-st is precursor, whereas the 2-nd – posterior to the *Meissner* expel) indicate an existence of <u>*3 distinctive temperatures at N/S transition*</u>. Crucial role of these specific temperatures for understanding of the real nature of the whole *SC*-phenomenon is discussed in the paper. In particular, the possible relation between these temperatures and the "*singlet*"/ "*triplet*" behavior of electron-pairs on the one hand and known *two 'ideal' properties* of the *SC*-matter ($R=0$ & $B=0$) - the other, is pointed out and carefully argued.

___________________________________________________________________

## 1. Introduction

There are no any doubts in the large progress in high-$T_c$ superconductivity (**HTSC**) since its discovery in 1986 [1]. Nevertheless, the nature of superconductivity (**SC**) in these materials is not clear yet [2]. Seems, one of the reasons is lack of test-methods for high-resolution non-destructive study of normal-to-superconductive (**N/S**) phase transition in *HTSC* at conditions close to the start of formation of the *SC*-state. Due to lack of resolution of test-methods near the transition (especially at study tiny, plate-like objects, such as clean *SC*-materials are with a small signal) we still deal with open questions in a basic superconductivity (related with low-$T_c$ *SC*-materials too - **LTSC**): details of problems see below. For answers one needs to study a beginning of formation of the *SC*-state. In this connection, the <u>*Meissner*-expel precursor</u> *'Para-Magnetic'* (**PM**) peculiarity (a fine *'unti-meissner'* effect noticed at very start of transition - detected first in *LTSC* <u>tin</u> granules [3-4] (Fig.1) and confirmed then in *HTSC* <u>YBaCuO</u> grains by Gantmakher et al. [5] - Fig.2) is an empirical evidence of unknown inter-phase physics near $T_c$, to be studied. Next evidence of the undisclosed physics around the $T_c$ one may find again in <u>tin</u>: namely, on heat-capacity *vs.* temperature curves, detected more than half century ago by Corak et al. [6-7] (Fig.3). We mean a little peculiarity noticed before the specific-heat's well-known jump. And it is strange enough, why such a key effect passed unnoticed so far. These are the reasons why the progress in our knowledge on the real nature of the superconductive phenomenon is impossible without high-resolution, non-destructive study of inter-phase physics between the normal and superconductive states in so tangled fluctuation region – around the $T_c$, which relates evidently with the complicated behavior of electron-pairs at very beginnings of their formation.

And so, for a long time there was need to create a test method below [8-9], which could move ahead studies of the *'paramagnetic'* effect also in thin-film *SC*-objects - capable of detecting other nuances of *SC*-phase transition too. Armed with such a tool our combined group began to analyze a fluctuation region and could actually reveal the *'PM'* effect also in YBaCuO film with expected huge signal-to-noise ratio (Fig.4). Then, the method enabled to open also a little absorption of a power by the YBaCuO film upon its *N/S* transition, with a peak located after the onset of *Meissner* expel. Extracted from the fluctuation region these two effects point out to the existence of *3 distinctive temperatures at N/S transition*. Important role of these specific temperatures for true interpretation of the nature of *SC*-phenomenon is the main goal of this paper. We discus relation between these temperatures & the *"singlet"* and/or *"triplet"* behavior of electron pairs (Cooper pairs) on the one hand and the known 2 *'ideal'* properties of the *SC*-matter ($R=0$ & $B=0$) - the other.

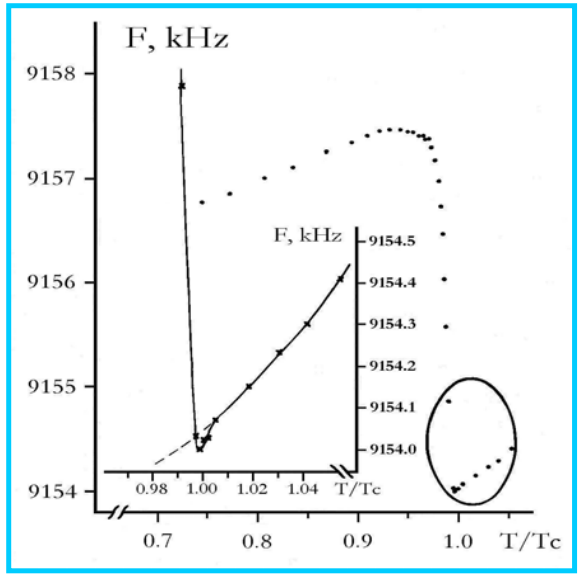

**Fig.1.** *'Paramagnetic'* effect detected upon transition to the *SC*-state of almost identical tin grains of ~ **5 µm** (± 0.5 µm) in diameter [3-4], *registered by a less sensitive solenoid-coil technique*. **Inset:** enlarged view of the effect. Broken line is device's temperature dependence.

## 2. Statement of the problem

It is known, that superconductor is a double *'ideal'* material, since it becomes an *'ideal conductor'* and gets properties of ideal diamagnetic below some temperature. The latter behaves also as the *'ideal'* conductor – reverse is not true. But, does the superconductive material obtain such properties at the same moment (temperature)? And why transitions of a different nature (1-st is assumed to be connected with electron-pairing & zeroing of the pair momentum and the 2-nd - with pair condensation (due to collection of enough *"singlet"* pairs - arguments see below)) should occur at the same temperature/moment? And also, are there different nature effects, which occur at the same moment in a Nature? Such basic questions arise since the said *'paramagnetic'* effect was detected in *micron*-size tin (*Sn*) grains [3] ($T_c\sim$3.72K - Fig.1), indicating the real physical onset of the *Meissner* expel. However, in a central press the effect was reported only a decade later [4] – after its obligatory check-up in many tin samples and in other *SC*-materials (including, in YBaCuO composition *HTSC* film - Fig. 4).

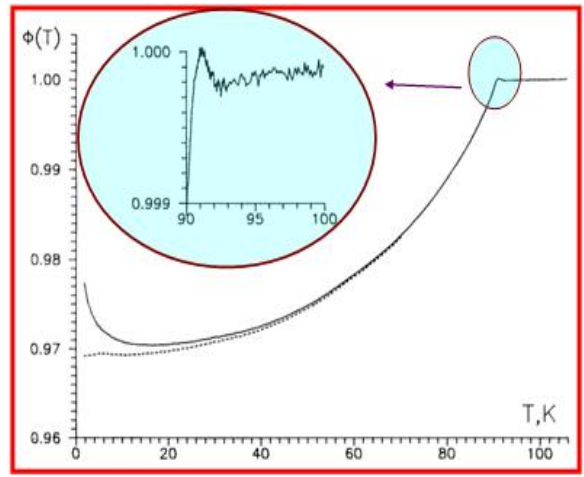

**Fig.2.** Superconductive transition of the ultra-fine YBaCuO powder with an average grain's size **2r** = **1920 Å** (± 40 Å) [5].
**Inset:** enlarged view of the *'paramagnetic'* effect.

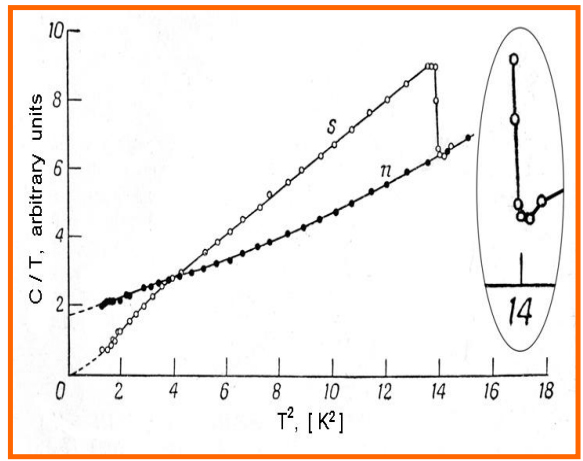

**Fig.3.** Heat-capacity *vs.* temperature curves detected in tin (**Sn**) [6-7]. **Inset:** enlarged view of the effect noticed before the known specific-heat jump. "*s*" corresponds to the SC-state, while "*n*" – to normal (superconductivity is suppressed by the magnetic field).

The *'PM'* effect precedes *Meissner* expel (known for 'helium' *SC*s since 1933 [10]) & noticeably corrects the shape of *N/S* phase transition curve. The origin of above questions relates also with the *'preceding'* effect, opened in percolating YBaCuO (in ceramics [1] and films [11] with a granular structure of the material). According to it *'resistive'* transition ends before the start of *Meissner* expel. To complete the picture note that it was seen also in BiCaSrCuO crystal [12], but doesn't attract a proper attention of authors - perhaps, due to lack of assurance in accordance among temperature-scales of conducted tests performed in different set-ups. Questions were especially deepened when a *'diamagnetic activity'* was revealed in LaSrCuO film by Iguchi et al. [13], at temperatures much higher the transition temperature of a material established by an onset point of *Meissner* expel. A super sensitive scanning-SQUID microscope was used for those tests. Such a flux activity was interpreted by the authors as the effect <u>precursor</u> to the *Meissner* stat*e*.

There are other <u>*Meissner*-state precursor peculiarities</u> too. Let's stop also on details of the research conducted by Tonica Valla's group [14]. It shows that a "*pseudo-gap*" in the energy level of high-$T_c$ materials electronic spectrum is the result of electrons being bound into the Cooper pairs above the transition temperature to the *SC*-state, but unable to super-conduct, because pairs move incoherently. As to traditional *LTSC* materials, which act much closer to the absolute zero temperature, <u>*it is admitted presently* that the superconductivity in LTSC materials occurs as soon as electron pairs are formed</u>. However, "in the case of high-$T_c$, electrons, though paired, "*do not 'see' each other* above some temperature," Valla says, "so they can't establish 'phase coherence,' with all the pairs behaving as a 'collective'".

*However*, our study indicates that, apparently, such is the case for all types of superconductive materials.

Origin of a "*pseudo-gap*", along with the mechanism for forming the pairs necessary for superconductivity, has been one of the biggest mysteries scientists have been trying to understand about high-$T_c$ superconductors since their discovery. The material studied by the Valla's group was the first high-$T_c$ superconductor discovered (LaBaCuO). In spite of the fact that this material at the ratio 1(Ba)**:**8(Cu) is not a superconductor it has a similar energy signature – *including energy gap in its electronic spectrum* ("*pseudo-gap*") - as the other *HTSC* materials in their *SC*-states. Valla's group interprets this finding as an evidence that the electron pairs are formed first (as "*preformed pairs*") and phase coherence occurs later, at some lower temperature (transition temperature), when thermal fluctuations of the phase are suppressed enough to cause superconductivity. Valla's research shows "that a "*pseudo-gap*" is caused by the same interactions that are responsible for superconductivity − interactions that bind 2 electrons into a Cooper pair". "In high-$T_c$ *SC* material, however, this pairing is only the first step," he says. "The phase transition is delayed, possibly - and ironically - because the pairing might be too strong. Figuratively speaking, a strong pairing produces "*small*" pairs with strongly fluctuating phases. Only by cooling the material to much lower temperatures do the phase fluctuations become suppressed. At that point, the phase becomes locked so the electron pairs can act coherently - and the system becomes a superconductor". *However, there is no indication on any microscopic physical mechanism for establishment of the phase coherence among the superconductive pairs in Valla's works*, as well as in the works of other researchers.

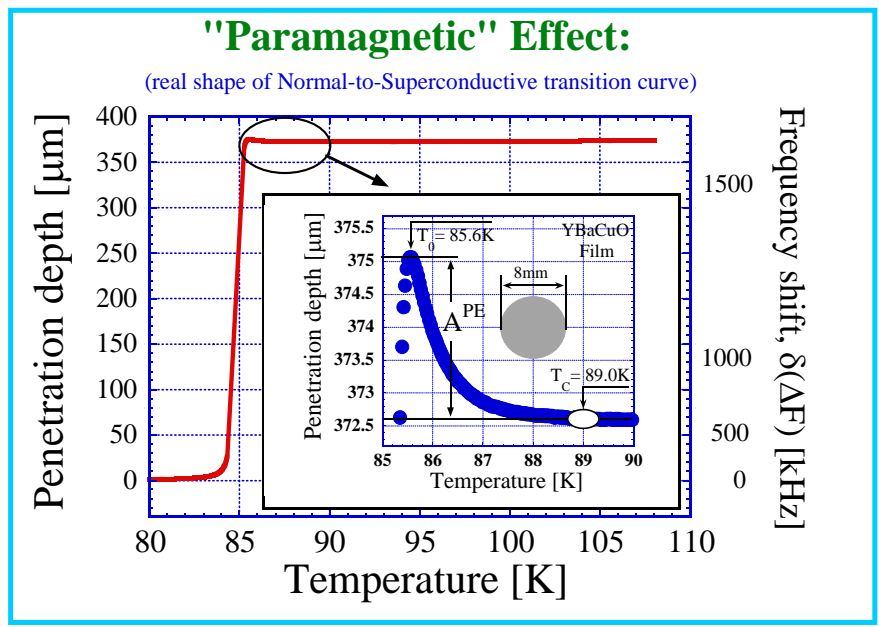

**Fig. 4.** Superconductive transition curve of the tested disk-shaped YBaCuO film [4] ($P_{osc.}$~ 4.8µW). **Inset:** − <u>enlarged view of a detected "paramagnetic" effect</u> (**PE**) at the beginnings of the N/S phase transition, which precedes the "*Meissner*" push-out. $A^{PE}$ is the effect's height.

*And so*, creation of *super charge-carriers* (Cooper-pairs) responsible for superconductivity seems starts at temperatures much exceeding an onset of the *Meissner* expel, which doesn't lie in frames of traditional theories. *Besides*, our test-method could detect a little increase of the penetration depth at beginnings of transition, upon cooling (in form of *'paramagnetic"* effect - Figs. 1, 4), which is also contrary to known conceptions. *And also*, nobody may ignore the said *'unti-jump'* effect, precursor to the jump of specific-heat (Fig.3), detected a long time ago but lost - due to lack of sensing of its importance. To all appearance, it is also caused by the physics, which is responsible for outwardly very similar *'paramagnetic'* effect. Anyway, this fine calorific effect is also beyond the traditional knowledge. *And besides*, to our knowledge, that is not yet detected in *HTSC* material.

*So*, the problem of electron-pairs' formation at start of *N/S* transition, and their behavior in the matter upon cooling still stays unclear, to be understood. *Frankly*, the whole physics of the *SC*-transition also still needs many-sided analysis in so tangled temperature-region. And so, conducting experiments near $T_c$ by non-destructive test-methods is still urgent problem. In this regard, *in-depth study of the shape of transition curve may serve as the effective "touchstone"*. The top goal for that may stand finding a relation between the shape of *N/S* transition curve and the said 2 *'ideal'* properties of the *SC*-matter.

*However*, established methods do not show enough resolution near the very start of transition. Besides, most of them are insensible for thin, plate-like, clean objects. On the other hand, seems also unjustified to use modern microwave methods for high-resolution tests near the $T_c$, since the electron pairs are destructed in such ultra-high frequency (*GHz*-range) testing fields. But, detection of so weakly expressed *'paramagnetic'* peculiarity even in a very small volume *HTSC* film by our *single-layer flat-coil-oscillator based test-method* (the **SFCO-***technique*) do encouraged our group to use it also in order to reveal and study other nuances of the *N/S* phase transition too, undisclosed so far by generally accepted methods.

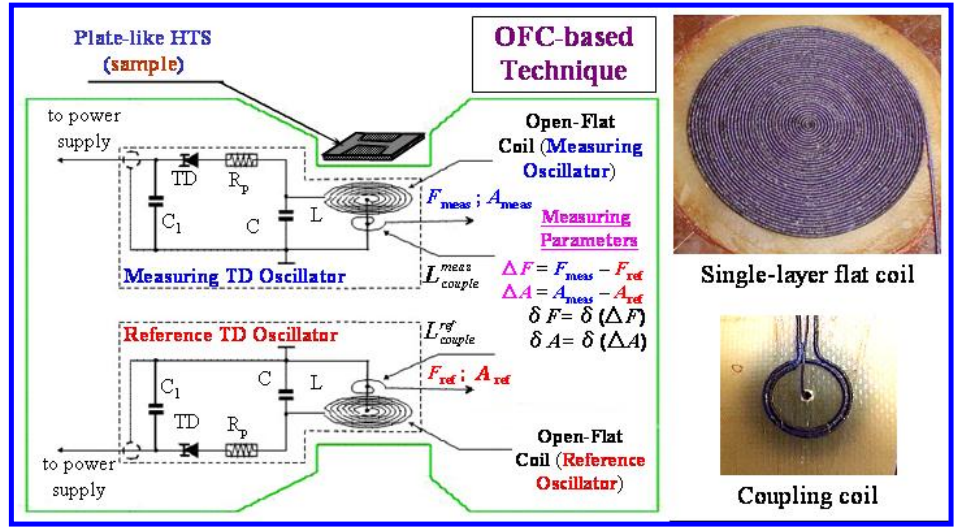

**Fig.5.** Schematics of the test-method based on a low-power tunnel diode (**TD**) oscillator with a single-layer open-flat coil [8-9].
**Side insets:** – the circular-shaped single-layer pick-up coil ($\Phi_{coil}$ ~ 8 mm), and the coupling coil ($\Phi_{couple}$~5mm).
($L$~1.5-2.0 μH;  C~20-40 pF;  C1~ 2500 pF;  $R_p$ ~ 150 Ω;  $F_{meas}$ ~ 23 MHz,  $\Delta F = F_{ref} - F_{meas}$ ~ 1.2 MHz,  $\delta F_{stability}$ ~ 2-3 Hz)

## 3. Flat-coil-oscillator technique and its advantages

A single-layer flat-coil-oscillator test-method [8-9] is a good tool for doing *MHz*-range measurements on thin, plate-like *HTSC* materials with small signals (Fig.5). It enables to *'notice'* fine peculiarities of transition in such objects. Due to flat shape of the coil, even a few pairs created in a flat *HTSC* lead to the strong deformation of a radio-frequency (**rf**) field configuration near the coil face at a *Meissner* expel. Similarly, even a little increase of a depth of skin layer of the material in normal state, due to a little fall in its conductivity caused by a leave of electrons from the Fermi surface at pairs' creation, may also lead to noticeable by the method distortion of a field configuration around the coil. Just such key features (the latter is even beyond the reach of SQUID technique) and high stability of the tunnel diode (**TD**) oscillator enabled to reach a record-high relative resolution in the present test-method, and detect fine effects. Among advantages of this method is its ability to reveal small amounts of energy-release ($\sim 10^{-10} - 10^{-9}$W) in a material caused by the movement of flux vortices created in *HTSC* material [15]. The method also enables detecting of record-small relative changes in *SC*-penetration depth, $\Delta\lambda/\lambda \sim 10^{-6}$ (abs. changes $\Delta\lambda$ ~1-3Å). Such abilities of this technique (as a *λ–test* method) is important, since, in terms of the amount and variety of data obtained from the tests, $\lambda$ is the most suitable detecting parameter[16-17]. Use of a flat coil in a *backward TD*-oscillator enabled to improve resolution of the tests in flat *HTSC* objects by 3-4 orders of a value [8-9]. That is so as replacement of a solenoid coil by the flat one made coil's filling factor maximal possible (~1) for flat objects, while its value for solenoid coil is about $10^{-4}$-$10^{-3}$. Both, the frequency & amplitude are used as testing parameters in a present method. The measuring effects are determined both by a distortion of a testing field's configuration near coil & by absorption of its power by the sample (due to various external or internal factors). *These finally lead to the changes of the TD-oscillator frequency or/and amplitude respectively.*

Note that flat coil's testing field can be distorted also by the low-$T_c$ superconductive film, during its transition. But, difference in distortion is more in the case of *HTSC* material. That is due to higher value of the skin depth in 'oxide' *SC*, in comparison with the ones in *LTSC*. This is because of much higher resistance of *HTSC* material in a normal state. So, a *MHz*-range testing field applied to the surface of *HTSC* (even several hundred $\mu$m-thick) passes through a sample leaving it from other face in a normal state, and almost completely (with an accuracy defined by the *Meissner* effect) is shielded by a sample in *SC*-state. While, much thinner *LTSC* film almost completely ejects the same waves both in normal and *SC* states - due to negligibly small value of a depth of skin layer of the material [18-19]. *And therefore*, our *SFCO*-method may be applied also for sensitive study of the phase transition of thin flat *LTSC* materials, if thickness of the material is less than a depth of its normal-state skin-layer.

### 4. Available data on the *'paramagnetic'* (PM) effect

Studies of the *'paramagnetic'* effect goes effective in plate-like *HTSC*-objects (in terms of the higher value of a signal-to-noise ratio), performed by the *SFCO*-method, optimally-suitable for flat geometry samples. That is not surprising, if take into account the above-said significant difference between *HTSC* and *LTSC* materials.

The *'paramagnetic'* effect is reproducible at cooling and heating of a sample, regardless of its form. It is seen both in 'oxide' and 'helium' superconductors. Initially, it was revealed in $d=5\mu m$ spherical tin grains (*I-type SC*) [3-4] (Fig.1). However, the effect was soon detected also in ultra-fine YBaCuO powder (*II-type SC*) with average grain's size $2r=1920Å$ (Fig.2) with much complicated atomic structure [5]. Both of these results are obtained by less-sensitive solenoid pick-up coil based methods with a signal-to-noise ratio at most ~10 units, although massive samples were tested. And therefore, results in granules are qualitative only. Presently, the effect was detected also in a disk-shaped YBaCuO film (Fig.4) [4] and in a YBaCuO ring (with an external diameter of the ring of about 3mm, 0.2mm wide and 0.2$\mu$m thick [20]).

Note, that the *'PM'* effect gradually disappears either the sample size is larger, or the temperature-stability is bad [3-4]. This is because of the effect averaging, due to the temperature and/or material inhomogeneity rise in a sample volume. In this connection, let's state conditions when the effect was observed in films. That is important for true understanding of its nature. So, the temperature-uniformity over a whole area of the film was maintained better than *5mK* at the time of phase-transition. Besides, empirically was chosen maximal cooling and heating rates, above of which the effect gradually disappears. The temperature-inhomogeneity increases along sample area at high rates, and the smoothing of the effect is observed, as a result. So, results related with the *'PM'* effect were obtained only in small-size samples & under conditions when the effect was independent on cooling & heating rates of a sample & the temperature-stability / material-homogeneity throughout the specimen was high enough. To the best of our knowledge, there are *no similar results* (like by a nature of the phenomenon to *'PM'* effect) obtained on *LTSC* or *HTSC* films, crystals or granules, on massive ceramics or in wires, *performed by a highly sensitive SQUID technique* or by other generally accepted sensitive methods. If only, let's again remember an obvious similarity between the *'PM'* effect and wrongly unnoticed *'unti-jump'* effect, detected by the calorimetric method – compare figures 1-4.

### 5. Possible explanation of the *'paramagnetic'* (PM) phenomenon ( a simplified conception )

Initial ideas related with the origin of *'PM'* effect are discussed in [4]. At start of formation of the *SC*-state a few Cooper pairs created in a sample (as a result of electron pairing from the Fermi surface) are not enough (according to Valla, the pairs "do not 'see' each other" [14]) to shield notably testing *rf*-field near the pick-up coil. Although, it is too small in our technique ($P_{osc} < 5\mu W$) − as small as the backward *TD*-oscillator may provide. However, *going away even a little amount of electrons from the Fermi surface reduces* (a little, but nevertheless) *normal-state conductivity of a sample, which leads to increase in a depth of the skin layer*. There are also other reasons for skin depth's even greater rise at a further cooling of the sample - before the *Meissner* expel. Namely, scattering of electrons on newly created *SC*-pairs & next created small size *SC*-domains. These all are important attributes of the fluctuation region (until powerful push-out of a testing field by the pairs) and more or less contribute to the *'PM'* effect. As the first approximation, these factors are responsible, and finally determine so much complicated shape of the *N/S* transition curve (see Figs. 1-2, 4).

Thus, at the start of *SC* phase transition first the pick-up coil testing field should slightly enter into the sample, resulting in some drop of the oscillator frequency $F_{meas}$ − this is why the effect is called *'paramagnetic'*. For more correctness note, that $\Delta F = F_{ref} - F_{meas}$ in Fig.4. Besides, frequency, $F_{ref}$, of the reference *TD*-oscillator (used for compensation of test-device temperature-dependence [8-9]) is independent of the effect, so, it remains constant during our measurements (Fig.5). The said decrease in a measuring oscillator frequency at very start of the phase transition, $\delta(\Delta F(T))$, is reasonable to connect with the temperature dependence of normal-state skin depth $\delta(T)$ (increase in $\delta$, caused by the leave of electrons from the Fermi surface) by the very-well known formula (1) [19], which is based on the traditional "*two-fluid*" model of superconductivity, proposed by Gorter and Casimir [21]

$$\delta(T) = c / \sqrt{2\pi\sigma_1(T)\omega}, \qquad (1)$$

where $c$ is the speed of light, $\omega = 2\pi F_{meas}$ ($F_{meas}$ is the frequency of test-oscillator), $\sigma_1(T) = [e^2 n_n(T)/m] \cdot \tau$ is the normal conductivity (here $e$ and $m$ are the charge & effective mass of an electron), $\tau$ is the relaxation time, and $n_n$ is density of the "normal" charge carriers, which is connected with the total carrier density, $n$, in the said "*two-fluid*" model (a simple original model suggested for description of a behavior of *BCS*-superconductors) by the simple equation $n_n(T)=n \cdot (T/T_c)^\gamma$, where $T_c$ is the superconductive transition temperature.

At our fitting of the measured phase transition curve (Figs. 4, 6) we assumed that $\gamma = 2$ in above relationship for $n_n(T)$, since most results of measurements of the temperature-dependence of *HTSC* films' penetration depth, $\lambda$, are approximated using just this value for $\gamma$ (see, e.g., the works [1-13] in [22]). As a suitable fitting function for the beginning of a transition curve (the start part of '*PM*' effect), we used $\delta(T) \sim \delta(\Delta F(T)) = [A/((T-T_0)/T_c)] + B$, where $A$, $B$ & $T_0$ are fitting parameters. $B$ is not so large in many practical cases, and the correct value of $T_c$ one may determine empirically (by use of the method, illustrated in Figure 4 of the Ref. [4]).

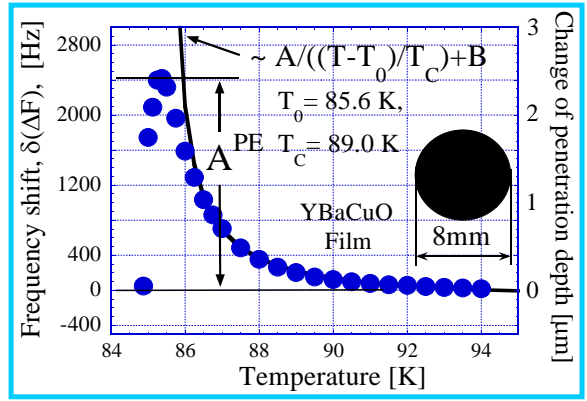

**Fig. 6.** The enlarged view of the "*paramagnetic*" effect (**PE**). $A^{PE}$ is the height of the effect and the bold line – the fitting by Eq.(1).

*At the subsequent cooling only, when the amount* (or capabilities) *of charge-carriers* (electron pairs) *becomes large enough to shield the testing rf-field* (or, by T. Valla [14], "electron pairs start act coherently, behaving as a 'collective'"), *a sharp rise in the measuring oscillator frequency is observed,* shown in the Figs. 1, 4. We mean a powerful push-out of the *MHZ*-range testing field from the surroundings of a single-layer flat shape pick-up coil by the sample (the well-known *Meissner* ejection).

## 6. To-day's interpretation of the *'paramagnetic'* (PM) phenomenon ( an improved conception )

Above, we have presented and discussed available data on *'paramagnetic'* effect. At that, we put attention on some details of test setups, permitting to reveal and study too much small in value, but significant in content this effect. We have mentioned also peculiarities of the samples, specified conditions under which the effect was detected, and gave some simple initial idea leading to its preliminary understanding - *not pretending, at that, on the finality of explanation of the phenomenon.*

Below, we discuss our to-day's interpretation of the *'paramagnetic'* effect, make conclusions and formulate far-seeing assumptions. Data collected on this effect [3-5, 20, 23-25], and on the mentioned above thermal [6] & absorption effects provide a solid basis in order to give a new interpretation of the physics of *N/S* phase transition (formation of electron pairs and possible changes in their behavior at further cooling of the material - especially at temperatures very close to the transition). *It may sound strangely enough, but seems the 'PM' effect should be considered as a decisive, not only for the correct understanding of the nature of HTSC materials, but also, for more complete and true understanding of the real nature of the whole superconductive phenomenon.*

### *Possibility of separation of the 'ideal conductive' state ($R=0$) from the superconductive one ($B=0$)*

*In fact*, the shape of transition curve (see Figs. 4, 6), corrected by the *'PM'* effect seems enables to separate an *'ideal conductive'* (the state without the resistance - **$R=0$**) and *superconductive* (ideal diamagnetic state - **$B=0$**) phase transitions. Besides, it allows connecting the experimentally measured shape of the phase transition curve with normal-state characteristics of the material.

*Really*, the shape of said function $[A/((T-T_0)/T_c)] + B$, which fits an initial part of the measured transition curve (onset of *'PM'* effect – see Fig.6), suggests that temperature-dependence of the "normal" charge-carrier density $n_n(T)$ (entering into Eq.(1) for skin depth, $\delta$, as the part of $\sigma_1(T)$), may not has a simple $n_n(T)=n \cdot (T/T_c)^\gamma$ form assumed by the traditional "*two-fluid*" model, but, more likely, it is much complicated. This dependence near the phase transition can be given, for example, by a relatively complex, but more realistic formula (2)

$$n_n(T) \cong n \cdot \{[T_c/(T_c - T_0)]^\gamma \cdot [(T-T_0)/T_c]^\gamma + n_{res}(T)/n\}, \qquad T_0 < T < T_c, \quad (2)$$

where $\gamma = 2$, $n_{res}(T_c) \cong 0$, and meanings of temperatures $T_c$ and $T_0$ are given and explained below.

Residual density of "normal" charge carriers in this formula, $n_{res}(T)$, is a slowly rising function of the temperature at a cooling. It is negligibly small for *LTSC* materials. However, as follows from the works [26-29], in *HTSC* it becomes evident (noticeable) in many important practical applications of these materials - due to presence of a "normal" fluid even at the absolute zero temperature. In order to take into account, and explain its appearance, a "*three-fluid*" model was proposed in Refs. [26-27] based on the concept of "*non-pairing*" residual charge carriers − $n_{res}^0 \equiv n_{res}(T=0) \cong const(T)$.

Temperature-dependence of "normal" charge-carrier density, $n_n(T)$, by Eq.(2) means that *"normal" electrons* starting from the $T_c \sim 89.0K$ (see Fig.4) gradually finish fulfilling of their physical 'mission' in a narrow region $T_c \rightarrow T_0$ (where $T_0 \sim 85.6K$ is a temperature at the peak of *PM'* effect) & their density notably drops as temperature of the material approaches $T_0$ (since $n_n(T_0) \cong n_{res}(T_0)$ - according to Eq.(2)). So, totality of pairs with a density $n_s(T)$ advances in formation (in a first approx.) **when temperature approaches $T_0$**. Apparently, starting from this moment *SC* material begins to possess an unusual property. Namely, **a noticeable part of pairs** (among all ones created in a matter) **start to correlate each other**. So, already at these temperatures a physical basis of the formation of a Bose-condensate of electron pairs is laid, **which**, *possibly*, **may come out from nulling of "spins" of those pairs, at $T_0$ approach**. These happens along with nulling of the momentum of all pairs - started for created first pairs since temperature has been dropped down $T_c$, notifying start of transition of the matter into '*ideal conductive*' state). We believe, that collection of some (minimal necessary) amount of pairs in a zero-spin ("*singlet*") state, as $T_0$ is approached (in a sample, where the pairs are present both in "*singlet*" & "*triplet*" states at any non-zero temperature below $T_c$) is just the reason leading to start of *superconductive* (ideal diamagnetic) transition in a matter. Besides, we think, at $T_0$ approach the matter could already finish its last efforts leading to the establishment of an '*ideal conductive*' state - owing to enough ideally conducting pairs created in a matter (particles with a zero momentum, so, with an infinitely large de Broglie wavelength, due to which the pairs can move in a matter '*ignoring*' defects, impurities and the thermal motion of the crystalline structure).

These are in a good agreement with data presented in Fig.7, where curves are shown for YBaCuO composition *HTSC* bridge (0.2mm wide, 4mm long, made of a 0.2μm thick film, patterned by the chemical etching method from the film identical to the one tested for figure 4), measured simultaneously by the standard 4-probe test-method (the *resistance*) & by our new *SFCO*-technique (*diamagnetic ejection*). As is seen, ~80% of a '*resistive*' transition ends before the start of a diamagnetic ejection. In other words, *bridge's transition into Meissner state starts when it is almost in R=0 state*. This is display of the said "*preceding*" effect in a tested bridge [11, 30]). A little difference (∼0.7K) between the start of *Meissner* expel & nulling of the bridge's resistance at our present tests is caused, probably, by the worsening of material properties along the bridge's edges during its etching. Better coincidence might be expected if more clean and homogeneous objects will be tested. *That is, for correct comparison of the relative positions of the 'resistive' & 'diamagnetic' transitions one needs to conduct tests on the single-crystalline HTSC or/and LTSC film-structures or crystals*. But, that is difficult to do, due to weak signals expected to detect from such too small-volume objects. Although, due to our recent estimations [31] the task is quite feasible for the improved *SFCO*-technique with about *1mm* in diameter, or smaller, pick-up coils (in terms of the higher spatial-resolution of tests). Thai is what we plan to do in our further experiments.

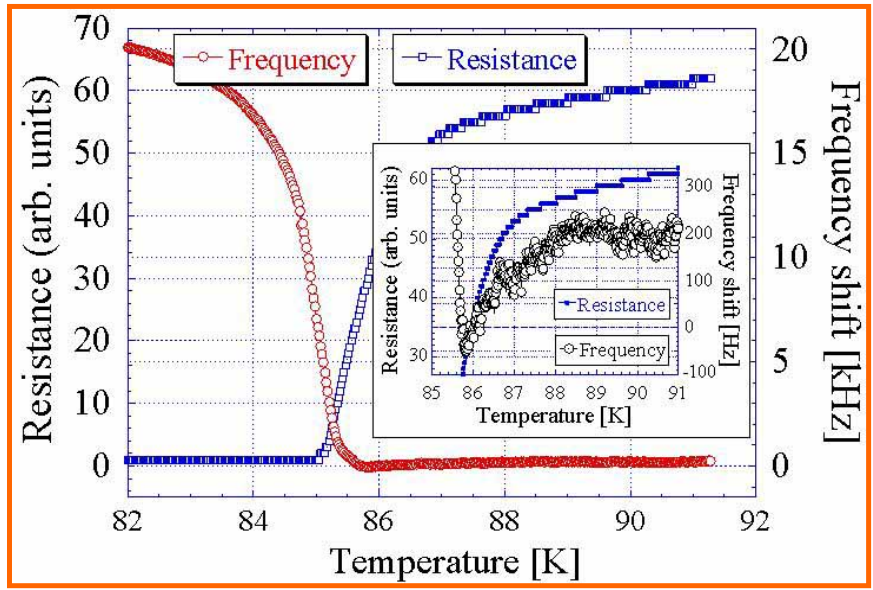

**Fig.7.** Normal-to-Superconductive transition curves of the YBaCuO-bridge detected simultaneously by two independent methods:
□ - 4-probe method (*resistance*), ○ - *SFCO*-method (*Meissner ejection*) shown in Fig.4, based on the frequency-shift of *TD*-oscillator.
**Inset:** the enlarged view of the curves just above the *Meissner* ejection.

Further cooling of the material below $T_0$ leads, most likely, to changes in a behavior of the superconductor described by the well-known *BCS*-theory (*working well in the case of "helium"* (simple) *superconductors* [32]) on the basis of still existing in a matter a few "normal" carriers (falling by the law $n_n(T) \cong n_{res}^0 + [n_{res}(T_0) - n_{res}^0] \cdot (T/T_0)^\gamma$; $T<T_0$), and already formed in the *SC*-matter large amount of electron pairs with a density

$$n_s(T) = n - n_n(T) \cong$$
$$\cong n - n_{res}^0 - [n_{res}(T_0) - n_{res}^0] \cdot (T/T_0)^\gamma, \quad (T<T_0), \quad (3)$$

where the number of "*non-paired*" residual "normal" charge carriers $n_{res}^0 \neq 0$ for 'oxide' superconductors. In contrast, $n_{res}^0 \equiv 0$ for "*helium*" (simple) superconductors.

*Note, that such conclusions follow from the corrected shape of the phase transition curve (see Figs. 4, 6) & are entirely opposite to traditional conceptions*, according to which positions of the '*resistive*' and '*diamagnetic*' transitions practically coincide (at least for the *LTSC* it is admitted, that $T_0 \cong T_c$) & the number of superconductive pairs increases at cooling, according to the simple law $n_s(T) = n \cdot [1 - (T/T_0)^\gamma]$ [21] - starting with onset point of a diamagnetic (*Meissner*) expel $T_0$ (in our notations).

### 7. Consequences caused by *'Meissner'* precursor & posterior <u>Subtle Effects:</u> *their crucial role for true identification of the nature of the SC–phenomenon*

<u>*So*</u>, data collected on the '*paramagnetic*' effect give us solid arguments for the next <u>*fundamental statement:*</u> seems *this effect enables to separate 'ideal conductive' (**R=0**) and superconductive (**B=0**) phase transitions* (and related states). That is illustrated by the curves shown in Fig.7, measured for the tested YBaCuO thin-film bridge by '*resistive*' and '*magnetic*' methods simultaneously.

*In other words*, corrected by <u>the 'PM' effect</u> shape of transition curve <u>selects temperatures $T_c$ & $T_0$</u> (an onset & peak of the effect - Fig.4), <u>which seems are connected with above 2 "ideal" properties of the SC-matter</u>. To be exact, $T_c \sim 89.0K$ *points to the start of 'ideal conductive' transition* (start of formation of "*triplet*" and/or "*singlet*" electron pairs with a zero momentum), *while $T_0=85.6K$ indicates both the start of Meissner expel* (start of long-range phase coherence among electron pairs leading to their further final phase condensation: <u>*in other words*</u>, collection of enough "*singlet*" pairs leading to the final phase coherence) *and final establishment of the 'ideal conductive' state (**R=0**) in a superconductive matter*.

<u>*Besides*</u>, according to the Eq.(1) an initial part of the '*paramagnetic*' effect is connected with the important normal-state characteristics of the *SC*-mutter, such as $m$ (effective mass of an electron), $\tau$ (relaxation time of "normal" charge carriers) and $n_n$ (their density).

These are questions, which are under wide discussion presently (see [33] and Refs. there). The problem of the electron pairing above the *Meissner* expel appeared after 1986 only, when *HTSC* materials were discovered. <u>*And almost commonly is admitted now, that there is need to consider 2 processes for the HTSC*</u> - electron pairing & onset of phase coherence - separately and independently of each other [34]. *Superconductivity in HTSC requires both electron pairing & Cooper-pair condensation* (the later is also known as an onset of the long-range phase coherence among pairs). <u>*Otherwise, it is admitted now,*</u> *that in high-$T_c$ materials quasi-particles become paired above the Meissner ejection* (above $T_0$, starting with $T_c$ – according to our notation – see Figs. 4, 6, 8) *and start forming SC-condensate only at the $T_0$, while in low-$T_c$ superconductors the pairing and onset of the phase coherence happens at the same temperature* (at $T_0$).

<u>*However*</u>, detection of so fine '*paramagnetic*' effect also in low-$T_c$ superconductive material (<u>in tin (**Sn**) - see Fig.1</u>) provides a weighty argument to have an opposite opinion regarding *LTSC* materials. And so, ***we assume that, most likely, there are NO essential differences between high-$T_c$ & low-$T_c$ materials regarding the said 2 processes:*** *apparently, electron pairing and onset of the phase coherence are separate & independent even in LTSC*. <u>*Difference is in the temperature scale only*</u>. As is seen from the Fig.1, for the *LTSC* material the process runs in a narrow range (*10-30 mK*), while for *HTSC* the scale is much longer (e.g., for YBaCuO that is broader *3-4 K* – Fig.4). Possibly, just this is the main reason why separation of $T_c$ from $T_0$ for low-$T_c$ materials was so problematic & unrealizable so far. There is no doubt that the problem is yet open also due to lack of test-methods for non-destructive study of *SC* transition in super-clean (so, tiny) *SC*-objects with too small signals - especially at very start of transition, where even an excellent SQUID technique is incapable to sense anything in changes in a normal-state penetration depth (skin-depth).

*That is why an accurate checking of this problem in low-$T_c$ materials is an important task for fundamental superconductivity*. As a proper tool for such a study one may use a flat-coil-oscillator based our unique method (*SFCO*–technique). In order to get deep insight into the real nature of the whole superconductive phenomenon such a very fine experiment may stand quite crucial.

*In this regard, seems urgent also searches for the absorption in LTSC material – detected so far only in HTSC*. In the case of YBaCuO the effect is posterior to *Meissner* expel, with a peak at $T_{abs}$=**84.9K**, located close to the center of transition (Fig.8). The effect is detected by the *SFCO*-method. And so, our test-method could define also the <u>3-rd distinctive temperature</u> of *SC*-phase transition. The effect is explained phenomenologically in [35] taking into account Coulomb interactions in a solid-state plasma composed of normal ions (the atomic core), as well as normal conducting and superconductive electrons. According to author of the paper there is an unknown reason (and a respective <u>physical process</u>) for such a very little rise in energy of the superconductive matter at very beginnings of *N/S* phase transition, before one may observe the well-known gain in a free energy of the *Meissner* phase with respect to the normal one.

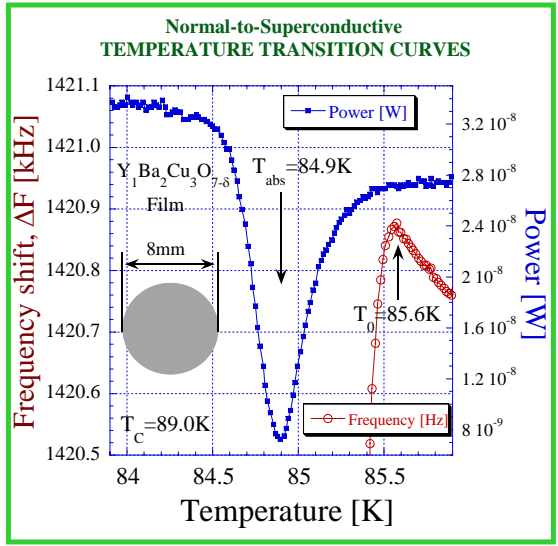

**Fig.8.** Absorption of a power by the YBaCuO-film with a clear peak at $T_{abs}$= 84.9K (the blue curve – *amplitude measurements*) positioned after the onset point of the *Meissner*-expel (the red curve – ***frequency measurements***), detected by a highly sensitive **SFCO**–technique.

Due to page limitation, we may not go into details of another fine effect (named *'unti-jump'* effect) shown in Fig.3 [6-7]. Precursor to the known jump of specific-heat, this thermal effect (and its relation to outwardly like *'paramagnetic'* effect) we discuss in [36]. This key effect passed unnoticed in due time, perhaps due to lack of feeling of its importance for basic superconductivity. In order to sense similarities & differences among high-$T_c$ and low-$T_c$ SC-materials we think, ***there is also need to repeat this calorific effect in a low-$T_c$ material, but with better resolution compared to that reached for tin*** (detected *'unti-jump'* effect is hardly seen in Fig.3) ***and search for this key phenomenon in high-$T_c$ materials***.

How much is true such an interpretation of physical processes determining the corrected (by *Meissner* expel precursor "*paramagnetic*" phenomenon and posterior to it *'absorption'* effect) shape of *N/S* transition curve may show additional multi-analysis in this topical area? The answer on fundamental question one may get conducting versatile tests on clean *LTSC* & *HTSC* samples by super-sensitive modern methods. But even now it is evident, that further studies in this area may essentially improve our understanding of the real role & behavior of electron pairs (Cooper pairs) in superconductive materials in so much tangled fluctuation temperature-region.

## 8. Conclusion and final remarks

First of all note that it was possible to extract from so tangled fluctuation region *Meissner*-expel precursor & posterior subtle effects only due to high potential of our technique based on a single-layer flat-coil-oscillator (the *SFCO*-method). Such an unusual planar coil forms the basis for creation of a *'magnetic-field'* probe enabling to formulate the new approach to surface probing based on substitution of the present solid-state, near-field probes by non-solid state, long-distance action ones [37]. Gap-size between such a probe and object may be larger by several orders of a value compared to what is the case in acting tunneling [38-39] and atomic-force microscopes [40]. After a proper perfection and adaptation of flat-coil probe to a design of acting probe microscopes one may get new microscope with non-contact far-ranging probe. It may enable to analyse *sub*-micron scale objects with a small local perturbance for the object. *Such microscopes may permit to distinguish magnetic regions from non-magnetic ones* (by the sign of a measured frequency shift)**,** *which is important for study coexistence of the superconductivity and magnetism* − detected at the moment in *sub*-micrometer scale objects only [41].

Such a probe one may use also to study the *'uniform superconductivity − to − non-uniform superconductivity' transition*, predicted in 1964 by Fulde & Ferrell and independently by Larkin & Ovchinnikov [42-43] (the so-called *FFLO*-state), and reported empirically by the U.S. combined group [44]. In such an unusual state (basically, different from the conventional *BCS*-case [32]), the magnetic field tries to polarize opposite spins of Cooper pairs. In response, the superconducting order parameter develops nodes in a real space, leading to the alternating *sub*-micron size regions of superconductive layers and spin-polarized magnetic walls. In this regard note that the lateral resolution reached in our method presently with about *10mm* size coil is around *1μm* [37]. And so, there is still need to improve the probe's lateral resolution by at least an order of a value to reach the limit enabling to detect reliably & study the said *FFLO*-state. *We expect that having an improved planar coil-based oscillator with ∼1mm size* [45] *or smaller pick-up coil, as an effective probing element, one may detect* (& further visualize) *the said node-structure in FFLO-state*.

### Acknowledgments

This study was supported by the **NFSAT** (*National Foundation of Science & Advanced Technologies*) & US **CRDF** (*Civilian Research & Development Foundation*) under **Grants # ISIPA 01-04** and **# UCEP 07/07**. The study was partially supported also by the state sources of the Republic of Armenia in frames of the task program on *'New Materials'* and *R&D* project # 301-0046.

Besides, author is grateful to Prof. V.F. Gantmakher for his comments on *'paramagnetic'* effect detected first in tin grains two decades ago, which stimulated further studies in *HTSC* films discussed. The author thanks also to Profs. M. Takeo, K. Funaki, T. Matsushita and Dr. T. Kiss for given chance to work and conduct research in Kyushu University, which promoted and accelerated assembling of experimental data on this problem (jointly published earlier), used here for discussion & motivation of new ideas. Author appreciates also discussions of test data & the problem as a whole with acad. D. Sedrakian, Profs. E. Sharoyan and Alex Gurevich. He also thanks to